\documentclass[12pt]{iopart}
\usepackage{epsfig}
\newcommand\hepph[1]{hep-ph/#1}

\newcommand\cpc[3]{{{\it Comput. Phys. Commun. }{\bf #1} (#2) #3}}

\newcommand\prl[3]{{{\it Phys. Rev. Lett. }{\bf #1} (#2) #3}}

\begin{document}

\vspace{-4cm}
\begin{flushright}
  UR--1589\\
  ER/40685/939\\
  hep-ph/9911536\\
  November 1999
\end{flushright}

\title{Vector Boson Transverse Momentum Distributions 
at the Tevatron}

\author{G. Corcella\footnote{Talk 
given at the UK Phenomenology Workshop on Collider Physics,
Durham, U.K., 19-24 September 1999.}}
\address{Department of Physics and Astronomy, University of Rochester,\\
Rochester, NY 14627, U.S.A.}
\author{M.H. Seymour}
\address{Rutherford Appleton Laboratory, Chilton,\\
Didcot, Oxfordshire.  OX11 0QX\@.  U.K.}

\begin{abstract}
We show vector boson transverse momentum distributions at the Tevatron,
obtained by running the HERWIG 
Monte Carlo event generator with matrix-element corrections.
We compare our results with some recent D\O\ and CDF data.
\end{abstract}
%-------------------------------------------------------------------
Vector boson production at hadron colliders is a fundamental process
to test 
Quantum Chromodynamics and the Standard Model of the electroweak interactions.
The lowest order processes $q\bar q'\to W$ and $q\bar q\to Z/\gamma^*$ are not 
sufficient to perform reliable phenomenological predictions, but the 
initial-state radiation has to be taken into account.
A possible way to deal with such multiple emissions consists in running a 
Monte Carlo event generator.
Standard Monte Carlo algorithms \cite{herwig} describe parton cascades in 
the soft/collinear approximation, with `dead zones' in the phase space which 
can be filled by the using of the exact first-order matrix element.

In \cite{csdy} we implemented matrix-element corrections to the HERWIG
simulation of Drell--Yan interactions: 
we filled the missing phase space using the exact ${\cal O}(
\alpha_S)$ matrix element (hard corrections)
and corrected the shower in the 
already-populated region using the exact amplitude for 
every hardest-so-far emission (soft corrections).
For $W$ production at the Tevatron, about
4\% of the events are generated in the dead zone, about half of which
are $q\bar q'\to Wg$ events.
Similar results hold for $Z$ production as well.

\begin{figure}
\centerline{\resizebox{0.49\textwidth}{!}{\includegraphics{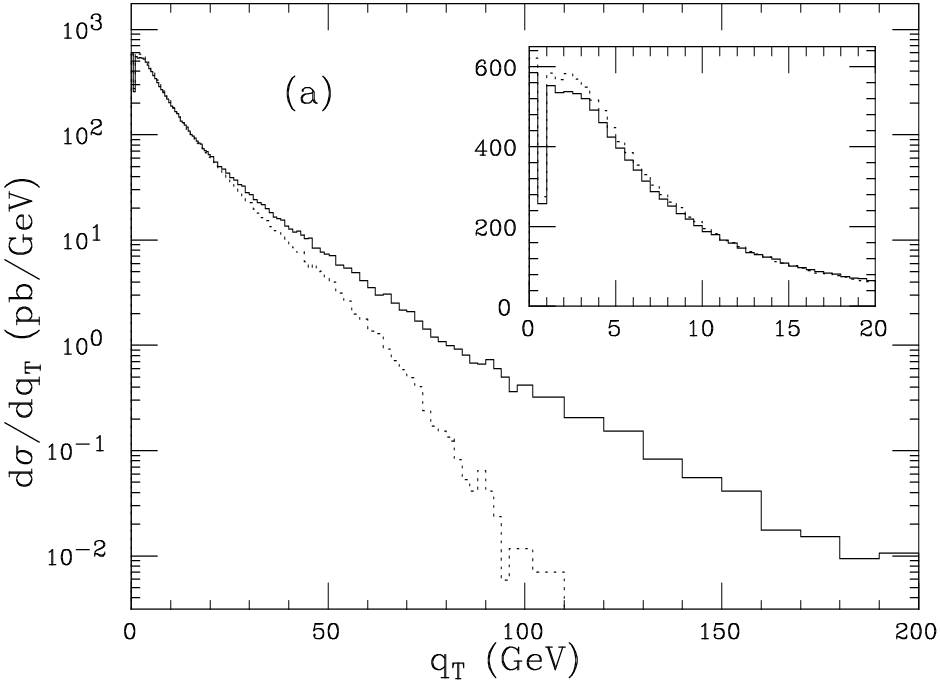}}%
\hfill%
\resizebox{0.49\textwidth}{!}{\includegraphics{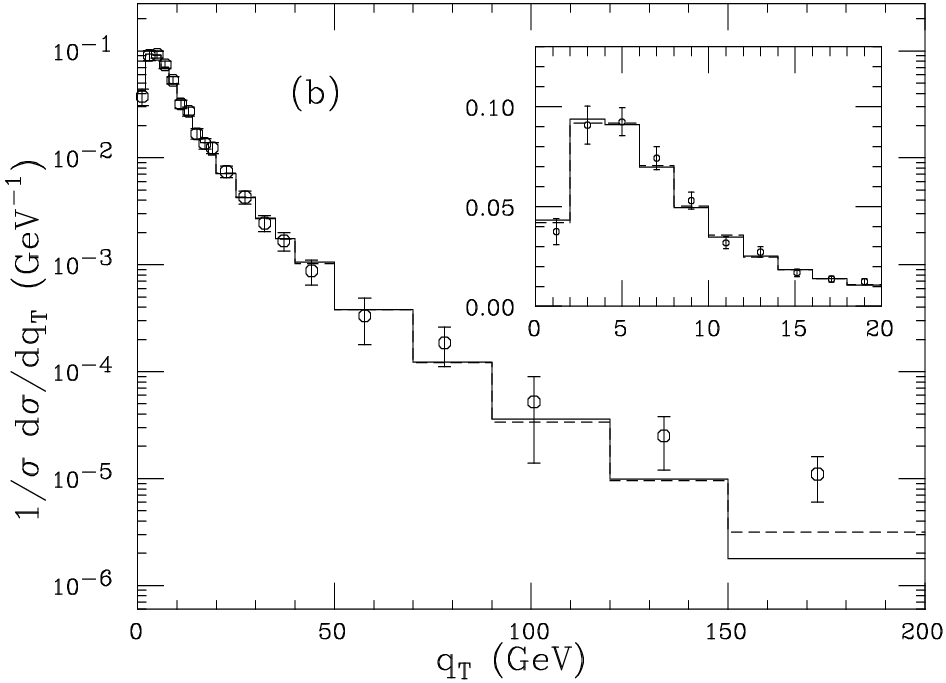}}}
  \caption{(a): $W$ transverse momentum distribution at the Tevatron according 
to HERWIG before (dotted line) and after matrix-element corrections (solid)
for $q_{T{\mathrm{int}}}=0$; (b): comparison of the D\O\ data with HERWIG 6.1
after detector corrections for $q_{T{\mathrm{int}}}=0$ (solid) and 1 GeV 
(dashed).}
\label{wqt} 
\end{figure}

An interesting observable to study is the vector boson transverse momentum 
$q_T$, which is the object of many theoretical and experimental analyses.
While in the parton shower approximation it has to be $q_T<m_{W,Z}$, after
matrix-element corrections a fraction of events with larger values of $q_T$ is 
to be expected.
In Fig.~\ref{wqt} (a), we plot the $W$ $q_T$ spectrum at the Tevatron 
according to HERWIG 5.9, the latest public version, and to 
HERWIG 6.1, the new version including matrix-element corrections
to Drell--Yan processes.
HERWIG has the option to vary the intrinsic transverse momentum 
$q_{T{\mathrm{int}}}$ of the incoming 
partons, which is set to zero, its default value, in 
the distributions shown in Fig.~\ref{wqt} (a).
We observe a remarkable impact of the corrections: after some
$q_T$ the 5.9 version does not give events anymore, while HERWIG 6.1
still has some events generated via the exact matrix element.
In Fig.~\ref{wqt} (b) we compare some recent D\O\ data 
\cite{d0} on $W$ production at the Tevatron with the HERWIG 6.1 results,
which we corrected in order to take the detector smearing effects into account.
We find good agreement overall; we also consider the option of 
$q_{T{\mathrm{int}}}=1$~GeV, but do not see any relevant 
effect after the detector corrections.

\begin{figure}
\centerline{\resizebox{0.49\textwidth}{!}{\includegraphics{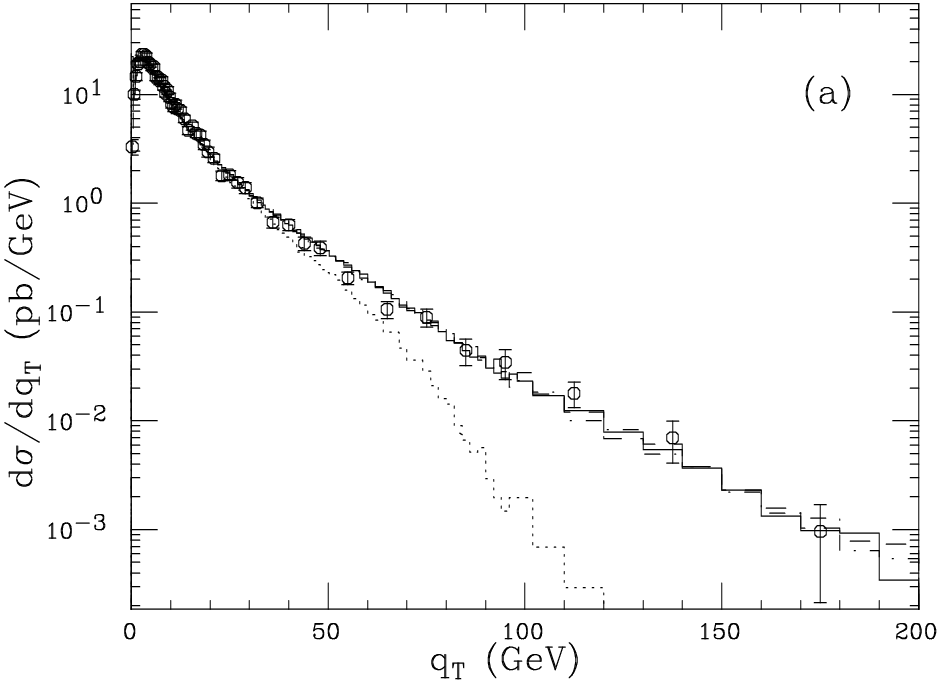}}%
\hfill%
\resizebox{0.49\textwidth}{!}{\includegraphics{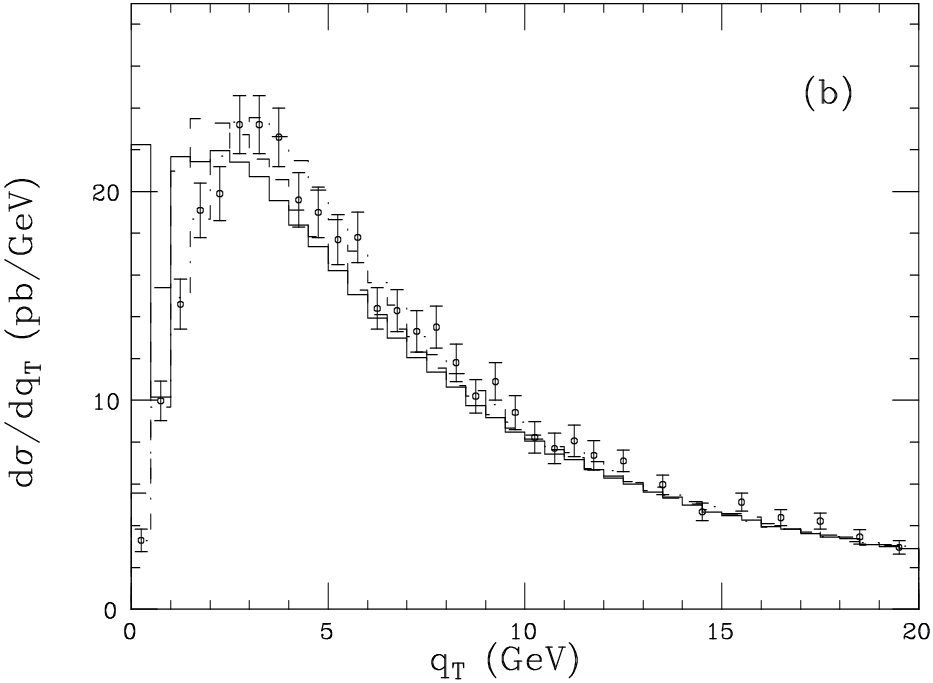}}}
  \caption{$Z$ transverse momentum distribution according to HERWIG 5.9
with zero intrinsic transverse momentum (dotted line) 
and according to HERWIG 6.1
with $q_{T{\mathrm{int}}}=0$ (solid),
1 GeV (dashed) and 2 GeV (dot-dashed), compared with the CDF data
over the whole spectrum (a) and for low $q_T$ values (b).}
\label{zqt} 
\end{figure}

\begin{figure}
\centerline{\resizebox{0.50\textwidth}{!}{\includegraphics{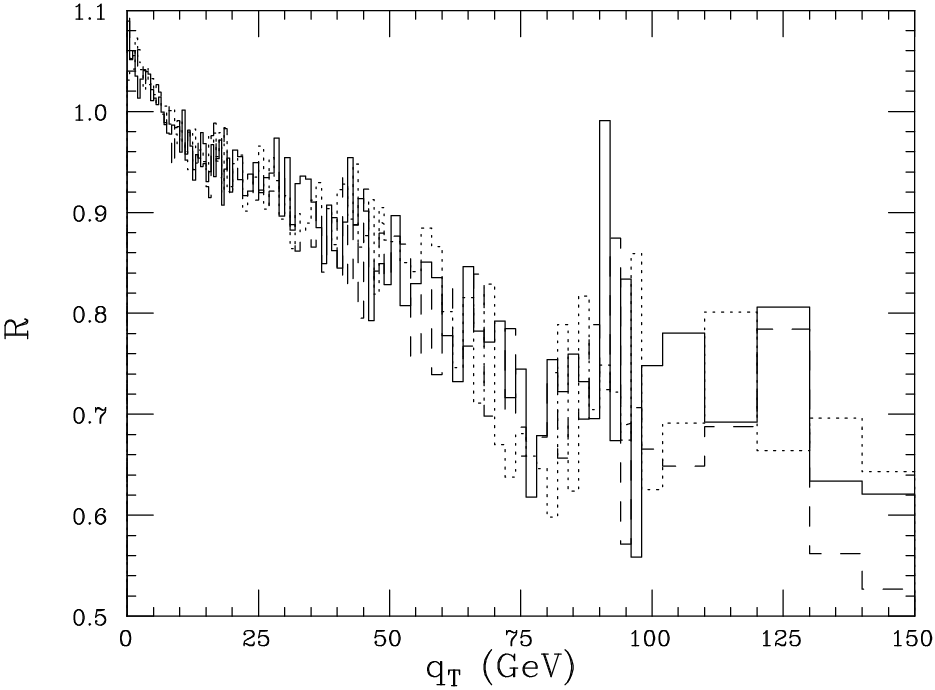}}}
  \caption{The ratio $R$ of the $W$ and $Z$ transverse momentum spectra,
running HERWIG 6.1, for $q_{T{\mathrm{int}}}=0$ (solid),
1 GeV (dashes) and 2 GeV (dotted).}
  \label{ratio}
\end{figure}

As far as $Z$ production is concerned, we have some preliminary CDF data 
\cite{cdf}, already corrected for detector effects, which
we compare with HERWIG 6.1 in Fig.~\ref{zqt}, where the options
$q_{T{\mathrm{int}}}$=0, 1 and 2 GeV are investigated. The agreement is 
acceptable and the role of the implemented matrix-element corrections is 
crucial in order to succeed in fitting in with the data for $q_T> 50$~GeV.
At very low $q_T$, the best fit is obtained by setting
$q_{T{\mathrm{int}}}=2$~GeV.
In Fig.~\ref{ratio}, we plot the ratio of the $W$ and the $Z$ transverse 
momentum spectra, both normalized to unity, for different values of 
$q_{T{\mathrm{int}}}$.  Although it can be seen from Fig.~\ref{zqt} (b) that
the $Z$ $q_T$ spectrum depends strongly on $q_{T{\mathrm{int}}}$ at low
$q_T$, the ratio of the $W$ and $Z$ spectra is insensitive to it.  This
is good news for the $W$ mass measurement in hadron collisions, as this
ratio is one of the main theory inputs that is needed.  A strong
dependence on unknown non-perturbative parameters like
$q_{T{\mathrm{int}}}$ could limit the accuracy of the $W$ mass
measurement at the Tevatron and, ultimately, at the LHC.

We have added matrix-element corrections to HERWIG's treatment of vector
boson production in hadron collisions.  They make an enormous difference
at high transverse momentum $q_T$, but little at low $q_T$.  Although
the dependence of the results on the non-perturbative intrinsic $q_T$
of partons in the proton ($q_{T{\mathrm{int}}}$) is quite strong at low $q_T$,
it is very similar in the $W$ and $Z$ cases, so that the ratio of the two
$q_T$ spectra is almost independent of $q_{T{\mathrm{int}}}$.

\end{document}